\documentclass[fleqn,12pt,twoside]{article}
\usepackage[headings]{espcrc1}
\usepackage{graphicx}
\usepackage[dvips]{color}
\usepackage{epsfig}

\readRCS
$Id: espcrc1.tex,v 1.2 2004/02/24 11:22:11 spepping Exp $
\def\PPNP{ Prog. Part. Nucl. Phys.}
\def\EPJC{{ Eur. Phys.J.} C}
\def\PLB{{ Phys. Lett.}  B}

\def\PRL{ Phys. Rev. Lett.}

\def\pTtrig{\mbox{$p^{\rm trig}_T$}}
\def\pTassoc{\mbox{$p^{\rm assoc}_T$}}

\def\dphi{\mbox{$\Delta\phi$}}

\def\Journal#1#2#3#4{{#1} {\bf #2}, #3 (#4)}

\title{Direct observation of dijets in central Au+Au 
       collisions with STAR}

\author{Dan Magestro\address{Department of Physics, The Ohio State University, \\
         Columbus, OH 43210} (for the STAR\thanks{For the full list of STAR authors and acknowledgments, see appendix Collaborations in this volume.} Collaboration)}
       
\begin{document}

\maketitle


\begin{abstract}
We present first results of azimuthal ($\Delta\phi$) correlations of high transverse
momentum ($p_T$) hadrons from the high-statistics Au+Au dataset collected at RHIC in 2004.  Previous correlations measurements observed the disappearance of back-to-back ($\Delta\phi\sim\pi$) correlations in central Au+Au collisions, supporting the emerging picture of parton energy loss in dense QCD matter.  In the present analysis, increasing the $p_T$ thresholds reveals a narrow back-to-back dijet peak which emerges above the decreasing combinatoric background.  We study the possible medium modification of the dijet fragmentation function and compare to models, to place constraints on the nature and magnitude of in-medium energy loss.

\end{abstract}

\section{Introduction}

The study of high transverse momentum ($p_T$) hadron production in heavy ion collisions at RHIC provides an experimental probe of the QCD matter in the most dense stage of the collision \cite{jacobsWang}, where quark-gluon deconfinement is likely to occur.  In particular, two-hadron azimuthal ($\Delta\phi$) correlations allow the study of back-to-back, hard-scattered partons that propagate in the medium before fragmentating into jets of hadrons, thereby serving as a tomographic probe of the medium.  Past Quark Matter conferences have seen significant first STAR results presented on the nature of back-to-back jets in Au+Au collisions.  Shown at Quark Matter 2002, the disappearance of back-to-back jetlike correlations of in central Au+Au collisions \cite{starB2bPrl} strongly supported the picture of parton energy loss, first seen in the suppression of inclusive hadron spectra \cite{phenixSupp}. At Quark Matter 2004, removing a lower-$p_T$ cut in the back-to-back correlation analysis instead revealed an excess of low-$p_T$ hadrons \cite{starFuqiang}, in contrast to the previously observed suppression of high-$p_T$ particles.  Taken together, the two results might indicate that gluons radiated by energy-degraded partons subsequently fragment into lower-$p_T$ hadrons that are still correlated with the other jet.

The performance of the RHIC accelerator during datataking in 2004 and an upgrade to STAR's data acquisition capabilities allowed STAR to collect an order of magnitude more Au+Au events than previously collected, extending the $p_T$ range accessible.  Searching for the re-emergence of the back-to-back correlation with increasing parton energy, as well as shifting to a $p_T$ regime where calculable pQCD fragmentation dominates particle production  \cite{phenixScaling,starK0Lambda}, motivate the analysis presented here.  In keeping with tradition, at this Quark Matter we present preliminary first results of higher-$p_T$ azimuthal correlations, where we observe distinct and narrow, back-to-back dijet peaks in central Au+Au collisions at $\sqrt{s}=200$ GeV.  The systematic study of dijet production and their characteristics is the most direct probe of both the medium's effect on propagating partons and the properties of the medium itself.  Such studies may eventually provide an upper bound on the gluon density and consequently a lower bound the number of degrees of freedom \cite{MullerRajagopal}.

\section{Experimental results}

As in our initial studies of high-$p_T$ azimuthal correlations \cite{starB2bPrl}, jetlike correlations are identified on a statistical basis by selecting high-$p_T$ {\it trigger} particles and measuring the azimuthal distribution of {\it associated} particles $\bigl(\pTassoc\!<\!\pTtrig\bigr)$ relative to the trigger particle.  The STAR Experiment \cite{starNim} is well-suited for azimuthal correlations studies due to the full azimuthal ($2\pi$) coverage of its Time Projection Chamber (TPC).  This analysis is based on 15.4M minimum-bias and 4.4M central Au+Au collisions at
$\sqrt{s}=200$ GeV, combining the 2001 dataset with a subset of the high statistics dataset collected during the 2004 run.  10M d+Au minimum-bias events collected in 2003 are also included in the analysis.  Event and track selection are similar to previous STAR high-$p_T$ studies \cite{starB2bPrl,starSupp1}. Charged primary tracks have $|\eta|<1.0$, with at least 20 (of 45 possible) space points in the STAR TPC.

\begin{figure}[t]
  \begin{center}
    \begin{minipage}{0.55\linewidth}
      \includegraphics[width=\linewidth]{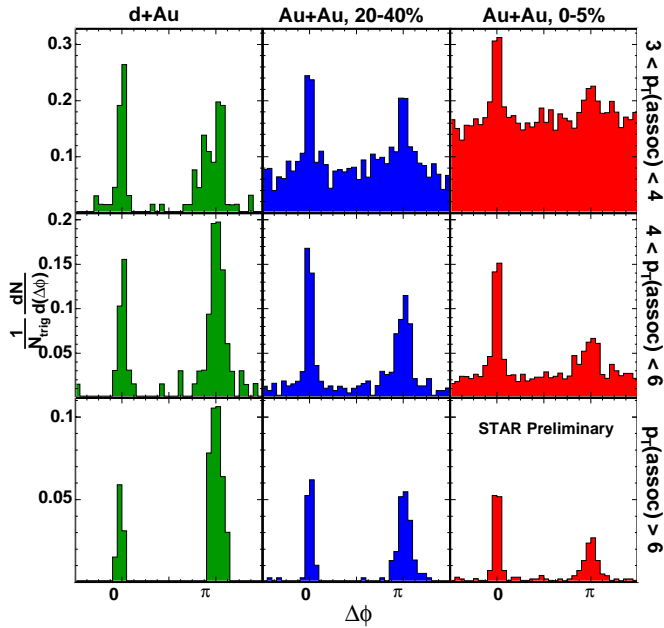}
    \end{minipage}\hfill
    \begin{minipage}{0.40\linewidth}
      \caption{Azimuthal correlation histograms of high-$p_T$ charged hadrons for $8 < \pTtrig < 15$ GeV/c, for minimum-bias d+Au, 20-40\% Au+Au and 0-5\% Au+Au events. $\pTassoc$ increases from top to bottom as indicated.  (STAR Preliminary)\label{figure1}}
    \end{minipage}
  \end{center}
\end{figure}

Figure \ref{figure1} shows azimuthal
correlations normalized per trigger particle with $8<\pTtrig<15$ GeV/c for mid-central and central Au+Au collisions, as well as d+Au collisions.  \pTassoc\ range increases from top to bottom in each
figure. The combinatoric background is not subtracted, is larger for more central collisions and decreases rapidly as
$\pTassoc$ is raised.  For the bottom panels ($\pTassoc > 6$ GeV/c) the background is nearly zero even in central collisions, simplifying the study of the jet peaks and reducing the systematic uncertainties in the underlying event subtraction.  Near-side ($\dphi\sim0$) peaks are seen in all panels and are similar in height for different systems.  Narrow away-side peaks are also visible in all panels, in contrast to published correlation results for lower \pTtrig ranges \cite{starB2bPrl}.  The away-side yield decreases from d+Au to central Au+Au collisions, following both the centrality dependence of the away-side yield at lower $p_T$ \cite{starB2bPrl} and the centrality dependence of the inclusive spectra \cite{starSupp1}.  We have also observed that the widths of the away-side peaks are consistent within errors among different system sizes.

\begin{figure}[t]
  \begin{center}
  {\includegraphics*[width=4.5in]{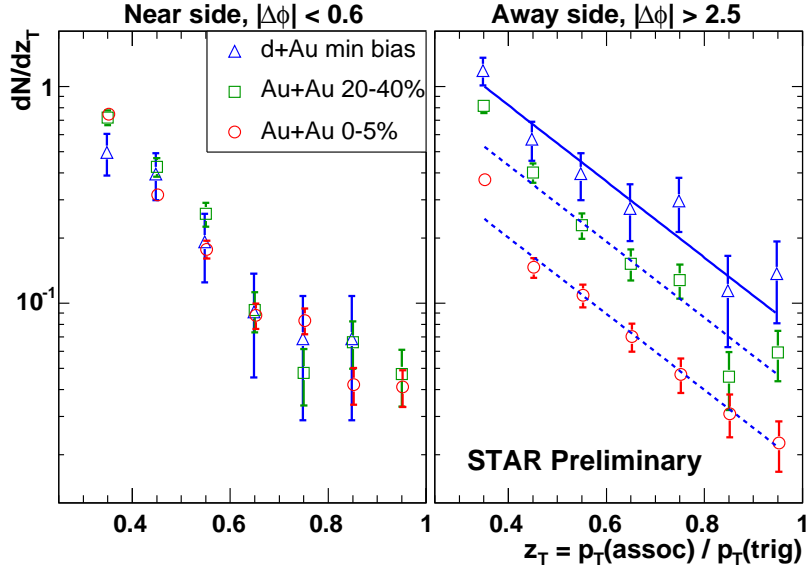}}
  \caption{(Upper panels) Transverse momentum distribution of near- and away-side charged hadrons in d+Au and Au+Au
  collisions at $\sqrt{s_{NN}}$=200 GeV, expressed in terms of the effective fragmentation variable $z_T$, the fraction of \pTtrig\ carried by the associated hadron ($8 < \pTtrig < 15$ GeV/c).  The solid line is an exponential fit to the $z_T$ distribution for d+Au; the dashed lines are scaled by a factor 0.54 (0.25) for  20-40\% (0-5\%) Au+Au.  (STAR Preliminary)}\label{figure2}
\end{center}
\end{figure} 

Figure \ref{figure2} shows the {\it hadron-triggered} fragmentation function \cite{xinNianFF} for both near-side and back-to-back hadrons for three systems with $8 < \pTtrig < 15$ GeV/c, where $z_T = p^{\rm assoc}_T / p^{\rm trig}_T$ provides an experimental handle on the parton's fragmentation without {\it a priori} knowledge of the parton energy.  The near-side $z_T$ distributions show little dependence on system size, disagreeing qualitatively with predictions \cite{Majumder04} that near-side energy loss would cause a factor 2-3 increase in associated hadron production due to changing trigger bias.  

The away-side $z_T$ distributions have very similar shapes for the three systems, as shown by an exponential fit to d+Au that is scaled to match the suppressed Au+Au $z_T$ distributions.  The suppression factor for central Au+Au collisions is $\sim 0.25$, quantitatively similar to suppression of inclusive spectra at similar $p_T$ \cite{phenixSupp}.  The similarity of the $z_T$ shapes in different systems, corresponding to a flat ratio of hadron-triggered fragmentation functions, is qualitatively consistent with calculations for medium-modified fragmentation due to energy loss \cite{xinNianFF}, although the data are 20-30\% below the predictions.  

\section{Discussion}

The similarity of the azimuthal width and suppressed spectral shape of the back-to-back associated hadrons in this analysis perhaps indicates that a fraction of parent partons have not interacted substantially with the medium.  In addition, the increasing suppression of away-side jets with increasing system size supports the picture that the observed dijets have short in-medium path lengths, perhaps oriented tangentially to the collision center.  A geometrical calculation \cite{daineseEtAl} of hard-scattering centers showed that collisions where both path lengths in the medium are less than $\sim 2.5$ fm are rare, indicating that the approx.~$25\%$ ``surviving" dijets likely have longer path lengths.  The non-interacting dijet scenario is a natural component of different energy loss formalisms \cite{xinNianFF,salgadoWiedemann} that contain path length-dependent terms for non-interacting partons, and quantifying the fraction of partons that survive would have direct consequences on the nature ({\it e.g.} collisional vs. radiative) and magnitude ({\it e.g.} gluon density) of parton energy loss.   However, recent calculations in a novel opacity approach to hadron correlations \cite{vitev05} show that the fragmentation of radiated gluons should cause an azimuthal broadening of the away-side jet for $p_T$ similar to those studied here, which is not seen in the data.

The observation of a clear dijet signal in central Au+Au collisions with vanishing combinatoric background opens up the field of dijet tomography of QCD matter, in the $p_T$ regime where perturbative QCD calculations are possible.  In STAR, systematic measurements are underway to determine the fragmentation properties of jets and back-to-back dijets in $\sqrt{s_{NN}}=200$ GeV d+Au and Au+Au collisions with 
high $p_T$ hadron correlations, the first results of which were presented in this contribution.  A comparison to existing theoretical calculations indicates that further refinements of the description of parton energy loss in dense nuclear matter are needed.  We already have indications from the data that we are seeing perhaps the fraction of dijets that have not interacted substantially with the medium, providing further insight into the interaction between high-energy partons and high-density QCD matter.


\end{document}